\documentclass[12pt,preprint]{aastex}

\slugcomment{To appear in ApJ Letters}

\shorttitle{Imaging Spectroscopy of the SgrA* Region}
\shortauthors{Krabbe et al.}

\begin{document}

\title{Diffraction Limited Imaging Spectroscopy of the SgrA* Region
using OSIRIS, a new Keck Instrument}

\author{A. Krabbe, C. Iserlohe}
\affil{I. Physikalisches Institut, Universit\"at zu K\"oln,
50937    K\"oln, Germany}
\email{krabbe@ph1.uni-koeln.de}

\and

\author{J. E. Larkin, M. Barczys, M. McElwain, J. Weiss, S. A. Wright}
\affil{Division of Astronomy, University of California, Los Angeles,
CA, 90095-1562, USA}

\and

\author{A. Quirrenbach}
\affil{Leiden Observatory, P.O. Box 9513, NL-2300 RA Leiden, The Netherlands}



\begin{abstract}
We present diffraction limited spectroscopic observations of an
infrared flare associated with the radio source SgrA*.  These are the
first results obtained with OSIRIS, the new facility infrared imaging
spectrograph for the Keck Observatory operated with the laser guide
star adaptive optics system.  After subtracting the spectrum of
precursor emission at the location of Sgr A*, we find the flare has a
spectral index (F($\nu$) $\propto$ $\nu^{\alpha}$) of $\alpha
= -2.6 \pm 0.9$. If we do not subtract the precursor
light, then our spectral index is consistent with earlier observations
by Ghez et al. (2005).  All observations published so far suggest that
the spectral index is a function of the flare's K-band flux.

\end{abstract}


\section{Introduction}

The low luminosity of the Supermassive Black Hole (SBH) in the center
of our Galaxy is a standing puzzle and challenges our understanding of
the mechanisms of mass accretion as well as the physics close to the
event horizon.  Infrared variability of SgrA*, first reported by
\citet{Gena03} and \citet{Ghea04}, has become an important observable.
The timescales (few minutes) imply that emission arises
from the immediate environment outside of the SBH.  Due to the
accretion's favorable duty cycle, flares are not only fairly frequent,
but they are also readily observable with adaptive optics at 8-10m
class telescopes \citep{Clea05, Ghea05}.  The infrared variability is
closely linked to the X-ray variability detected a few years earlier
\citep{Baga01,Pora03,Gola03}.  Simultaneous infrared/X-ray
observations have demonstrated that the flares are related
multiwavelength phenomena \citep{Ecka04}.  The spectral index of the
flare and its possible variation from flare to flare and during a
single flare are important parameters that will allow us to determine the
emission process of the radiation \citep{Yuaa03,Yuaa04,Liua04}.  We
observed SgrA* with OSIRIS, a new Keck facility instrument, during its
commissioning time \citep{Lara03,Kraa04,Quia03,Weisa02}.  These data
are the first laser guide star (LGS) assisted spectra ever taken of
the Galactic Center region.

\section{Observations and Data reduction}

The OSIRIS (OH Suppressing InfraRed Imaging Spectrograph) instrument
is a new facility near infrared (NIR) Z- to K-band imaging
spectrograph designed for the Keck Observatory's Adaptive Optics (AO)
system.  It utilizes an array of micro lenses and a HAWAII-2 detector
(2048$\times$2048 pixels) to simultaneously obtain more than 1020
spectra over a rectangular field of view with about 16$\times$64
spatial positions.  Each spectrum in this mode covers about 1700
channels at a spectral resolution of R = 3700. OSIRIS achieves high
sensitivity with good instrumental throughputs, low backgrounds per AO
resolution element and high enough spectral resolution to work between
the night sky lines.  First light was achieved on February 22, 2005, and
commissioning will be completed in 2006.  A detailed account of OSIRIS
will be given by Larkin et al. (in preparation).

The Galactic Center was observed on April 29, 2005 at Keck II during
the first deployment of OSIRIS with the LGS AO system (Wizinowich et
al., in prep., van Dam et al., in prep.).  Two consecutive K-band
frames, each with 300 s integration time were obtained at an airmass
of 1.65 with SgrA* in the field of view (FOV).  Overhead between
the frames was about 37 s.  The frames were followed by an empty sky
field obtained immediately after the $2^{nd}$ frame.  The angular
scale of OSIRIS was set to 20 mas/pixel.

A dedicated OSIRIS data reduction pipeline has been produced, and was
used to identically reduce both exposures. After sky subtraction, the
individual spectra in each frame
were extracted from the raw frame, using a special map of the point
spread function of each lenslet at all wavelengths.  This allows the
removal of crosstalk from adjacent spectra and correct assignment of flux to
each field position over the 2-dimensional field.  Arc line
spectra are used to calibrate the wavelength scale of each field
point.  Atmospheric differential dispersion effects were also
corrected by tracing the peak emission of the stellar continuum
through the wavelength slices of the cube.  Telluric and
instrumental transmission, as well as foreground extinction to the
Galactic Center, were corrected in both data cubes by dividing all
spectra by the average spectrum of star S2.  All spectra were finally
multiplied by a black-body curve of T = 30000 K, representing the
approximate flux density of S2 which is assumed to be between spectral
classes O8V and B0V based on spectra from \citet{Ghea03}. Since we didn't
attempt to model the stellar absorption lines, e.g., Br$\gamma$ at
2.166 \micron, the final spectra do not contain valid information
about emission or absorption lines at those spectral positions and
were ignored in our analysis.

\section{Results}

Figure 1 has two panels, hereafter referred to as frame 1 and frame 
2, that display
the image produced by collapsing all of the
wavelength channels from 2.02 \micron\space through 2.38 \micron\space
for each of the two frames. Some of the stars are labeled
according to \citet{Eisa05}. Frame 2 was observed 0.3\arcsec\space
south with respect to Frame 1.  The angular resolution on the sky is
60 mas at a pixel scale of 20 mas.  The angular resolution achieved is
worse than the diffraction limit of 46 mas at 2.18 \micron, probably
due to the low telescope elevation.

The staggered edge along the sides of each data cube is standard for
OSIRIS and is due to the complex mapping of lenslets onto the
detector. Due to constraints in the commissioning schedule, a global
flat was not available for these observations. This means that
individual spatial locations have a well calibrated spectrum, but
based on comparisons of known stellar fluxes in the field, the
relative intensity of one spectrum to another is uncertain at the 20\%
level (reflected in the flux errors in Table 1). Star S2 has a well
determined flux and was used to flux calibrate both frames individually.

The flare at the position of SgrA* is apparent in frame 2.  It is
located 145$\pm5$ mas south and 22$\pm5$ mas west of S2 at the epoch
of the observations: MJD53489.624680.  Inspecting the identical
position in frame 1 reveals a weaker but still notable emission at the
location of Sgr A*, which we will term the precursor. In both frames,
the emission is close to other stars making a proper background
subtraction important and slightly difficult. The levels and spectra
of the backgrounds at
SgrA* and S2 are certainly not identical and had to be treated
individually. In the end the spectrum
of SgrA* was determined from frame 2 using a 5 pixel $\times$ 4 pixel
box centered on the position of the flare. The general background was
determined by measuring the average pixel value in the 22 pixels that
form a circumference around the 5$\times$4 aperture. The spectrum of
S2 was determined in an identical fashion from the same frame, in order
to reduce systematic errors that could result from using different
fractions of the point spread function or by measuring the background
in a different manner.  An identical procedure was applied to frame 1
to extract the precursor spectrum, except a 5$\times$5 pixel aperture
was used due to a fractional shift of the lenslet grid on the sky
between frames 2 and 1. Again, the size and geometry of the box was
identical to that used for the extraction of the S2 spectrum discussed
earlier. Assuming that the flare is unresolved in
our data, the
extraction regions for SgrA* and star S2 then cover the same
fraction of the PSF for each source.

The resulting spectra are shown in Figure 2.  The lowest spectrum
represents the precursor emission in frame 1.
The middle spectrum is the extracted spectrum of the
flare from frame 2, and has has been shifted vertically by one unit
for clarity. The upper spectrum is the difference between the spectrum in
frame 2 and the spectrum in frame 1, and we will refer to this as the
flare spectrum. Again for clarity, it has been shifted vertically by 4 units.
All three of the spectra have
been smoothed by a 30 pixel wide boxcar filter. The sky frame was
obtained after both frames 1 and 2 and was a poorer match for frame
1. This was particularly important in the spectral range between 2.04
\micron\space and 2.06 \micron\space and this region was masked out in
frame 1. Atmospheric OH lines, however, subtracted out well. All three
of the spectra were fit with a power law ($F(\lambda) \propto
\lambda^{m}$) and the resulting fits and slopes (m) are given on Figure
2 along with $1 \sigma$ errors. These slopes were converted into a
frequency power law index $\alpha$ (F$_{\nu} \propto
\nu^{\alpha}$). Table 1 shows all of the derived quantities including
the slope and spectral index. We note here
that the spectral index of the flare is quite red
($\alpha$=$-2.6\pm0.9$).

  From the spectral fits, the K-band flux of the flare and the precursor
were determined relative to S2 by extrapolating the power law fit from
(2.02 \micron, 2.38 \micron) to the full K-band.  We assumed that S2
has a magnitude of $m_{K} = 13.9$ mag (24 mJy) based on photometry from
\citet{Ghea03}. The K-band flux of the precursor emission at the
location of SgrA* in frame 1 then becomes 3.5 mJy, or 6.9 times
fainter than S2. The corresponding flux at the location of Sgr A* in
frame 2 is 9.6 mJy, or 2.5 times fainter than S2. This is an
increase in the K-band flux of SgrA* by a factor of 2.8 within 5.6 minutes
(exposure time plus overhead between frames).

\section{Discussion and Summary }

While frame 2 (Fig. 1b) obviously shows a flare at the position of
SgrA*, it is less clear what is at the same location in frame 1 (Fig
1a). Is it a true quiet state or a mild flare precursor? A faint
source at the location of SgrA* has been found by other observers,
including in high resolution H-band images by \citet{Eisa05} with the
NACO instrument. Photometric K-band data by Genzel et al (2003)
indicate that SgrA* can be more than 6 times fainter than
S2. \citet{Ecka04} report the lowest activity level 12 times fainter
than S2 corresponding to a K-band flux density of 2 mJy. Comparing
this factor with Table 1 suggests that the K-band emission in frame 1
is less than a factor of 2 brighter than the lowest activity level measured and
may indeed be dominated by an underlying constant source. The spectral
index of the source in frame 1 ($\alpha$=$2.7\pm1.3$) is also
consistent with a blackbody at a temperature around 3000 K. This
suggests the possibility that the emission at the location of SgrA*
during low activity may be dominated by a different mechanism than the
flare itself. One option is a stellar component either in the
background or at the very cusp of the stellar cluster within less than
a light day of SgrA*.  If this is the case, then these faint
stars are also orbiting in the immediate vicinity of SgrA*.

The temporal separation between the two frames is 337
seconds (exposure times of 300 sec with a gap of 37 seconds).  Within
this time or shorter, the K-band flux increased by more than a
magnitude to 9.6 mJy.  This peak flux compares very well with
\citet{Ecka04} and also with \citet{Ghea05} for the maximum
observed activity.  Our observations thus very likely record the
beginning of a flare with a typical level of activity. We also
conclude that the intrinsic K-band flux of the flare is probably best
represented by
the difference between the peak value of 9.6 mJy and the
lowest background of 2 mJy, yielding ($7.6\pm 34\%$) mJy. This value
has been used in Figure 3.

\citet{Eisa05} were first to report measurements of the spectral index
of the emission of SgrA* in the NIR. Their spectral indices $\alpha$
for the flux density F$_{\nu}$ lie in the range between -3.3 and -4.8,
with an averaged index of $\alpha_{E} = -4\pm1$ for fluxes $<2$ mJy.
Their value has been
obtained in a similar fashion to ours, in that two spectra of
different activity levels were subtracted from each other.  However,
their result is different from our value of $\alpha_{K} = -2.6\pm0.9$
(see Table 1 and Figure 3).

Figure 3 summarizes all published spectral indices for SgrA*'s K-band
flares as a function of their 2 \micron\space flux density.  In a
recent paper, \citet{Ghea05} report spectral indices based on
K$^{\prime}$ and L$^{\prime}$ imaging.  Their method of determining $\alpha$
is directly based on K$^{\prime}$-L$^{\prime}$ PSF fitting photometry.
However, different from our result and from \citet{Eisa05}, their
result does not account for flux from the precursor source.  Thus
their value of $\alpha$ = -0.5 has a different meaning than our flare
index. It is more correctly compared to the slope of the spectrum in
frame 2 which includes the local background and has an index of
$\alpha = -0.6\pm0.4$, consistent with the Ghez measurement. In
addition, their K-band flux densities are close to our results,
especially if we correct for the difference between K$^{\prime}$
and K bands. Using their spectral index value of $\alpha$ = -0.5 and
K$^{\prime}$ flux density of 7.2 mJy, we calculate a K-band flux
density of 9.2 mJy, again very comparable to our measurement
of 9.6 mJy for frame 2.

We thus confirm the findings of \citet{Ghea05} for the slope of the
flare plus local background within the 1 $\sigma$ errors.  It is
interesting to note that both spectral indices were obtained by very
different techniques and using different wavebands: K$^{\prime}$ and
L$^{\prime}$ versus K band only.  The spectral indices reported by
\citet{Eisa05} are clearly different but were also taken at a
significantly lower level of activity.  A possible interpretation of
the difference in spectral index as a spectral break between 2.4 and
3.8 \micron was first suggested by \citet{Ghea05}.  Our new data show
that the K-band spectrum has a smooth constant spectral slope.  The
spectral slope of our data is close to the Ghez et al. value, leaving
little room for a spectral break between 2.4 and 3.8 \micron. Instead,
taken at face value, our data imply a dependency of the spectral index
on the flare intensity, another suggestion first made by \citet{Ghea05}.
This same conclusion has also been reached by \citet{Gill06} based on near
infrared spectra taken with the SINFONI instrument at the VLT.

Theoretical models by \citet{Yuaa04} and \citet{Liua04} indicate that
both the infrared and the X-ray variability is probably due to hot
or relativistic gas accelerated at $\sim 10$ times the Schwarzschild
radius, $R_{S}$.  The synchrotron emission from the high-energy
electrons within this gas accounts for the observed X-ray
flares.  The models can also be tuned to explain the infrared flux and
its variability.  The models are flexible enough to produce spectral
indices between -2.0 and -3.0, consistent with measurements from
this paper (see Figure 4 of \citet{Yuaa03}). We note that both
\citet{Yuaa04} and \citet{Liua04} explicitly mention that varying
spectral indices may be a feature of such flares.  More spectroscopic
data at different flare intensity levels are needed to confirm such a
trend.  If the flare spectral index is a function of the flare
activity level, it favors a model in which stronger flares involve
more energetic electrons, making the infrared spectrum bluer. Such
a finding would considerably narrow the range of process parameters
describing the origin of the flare.

\acknowledgments

The authors would like to sincerely thank the dedicated members of the
OSIRIS engineering team: Ted Aliado, George Brims, John Canfield,
Thomas Gasaway, Chris Johnson, Evan Kress, David LaFreniere, Ken
Magnone, Nick Magnone, Juleen Moon, Gunnar Skulason, and Michael
Spencer. We would also like to thank the Keck Observatory Staff (CARA)
who were part of the OSIRIS team and helped with commissioning.
Special thanks goes to Cara staff members Sean Adkins, Paola Amico,
Randy Campbell, Al Conrad, Allan Honey, David Le Mignant, Marcos Van
Dam, and Peter Wizinowich.

The authors would like to acknowledge the support of the Keck
Observatory and the Keck Science Steering Committee. The design
and construction of OSIRIS were supported by a grant from the
California Association for Research in Astronomy, which owns and
operates the Keck Observatory. A major portion of the CARA funds were
provided by the TSIP program from the National Science
Foundation. This work also received early support from the National
Science Foundation Science and Technology Center for Adaptive Optics
(CfAO), managed by the University of California at Santa Cruz under
cooperative agreement No. AST-98-76783.

{\it Facilities:} \facility{Keck (OSIRIS)}


\clearpage

\begin{table}
\begin{center}
\caption{Derived Quantities\label{tbl-1}}
\vskip 1cm
\begin{tabular}{ccccc}
\tableline\tableline  & & Spectral slope m & & Spectral Index $\alpha$ \\
Target & K $[$mag$]$ & ($F(\lambda) \propto \lambda^{m}$) & F$_{\nu}$ [mJy$]$
\tablenotemark{c} & (F$_{\nu} \propto \nu^{\alpha}$) \\
\tableline Star S2 (both frames) &13.9\tablenotemark{a} & & 24 & \\
SgrA* (frame 1) &$16.0\pm0.2$\tablenotemark{b} &$-4.7\pm1.3$
&$3.5\pm20\%$ &$2.7\pm1.3$ \\
SgrA* (frame 2) &$14.9\pm0.2$\tablenotemark{b} &$-1.4\pm0.4$
&$9.6\pm20\%$ &$-0.6\pm0.4$ \\
SgrA* flare &$15.4\pm0.3$\tablenotemark{b} &$0.6\pm0.9$ &$6.1\pm34\%$
&$-2.6\pm0.9$ \\
\tableline
\end{tabular}
\tablenotetext{a}{\citet{Ghea03}. No errors were provided.}
\tablenotetext{b}{Fit extrapolated from $(2.02 \micron - 2.38 \micron)$ to
full K-band (1.95 \micron\space - 2.40 \micron)}
\tablenotetext{c}{A$_{K} = 2.8 mag$ extinction corrected, \citet{Gena03}}
\end{center}
\end{table}


\begin{figure}
\epsscale{.60}
\plotone{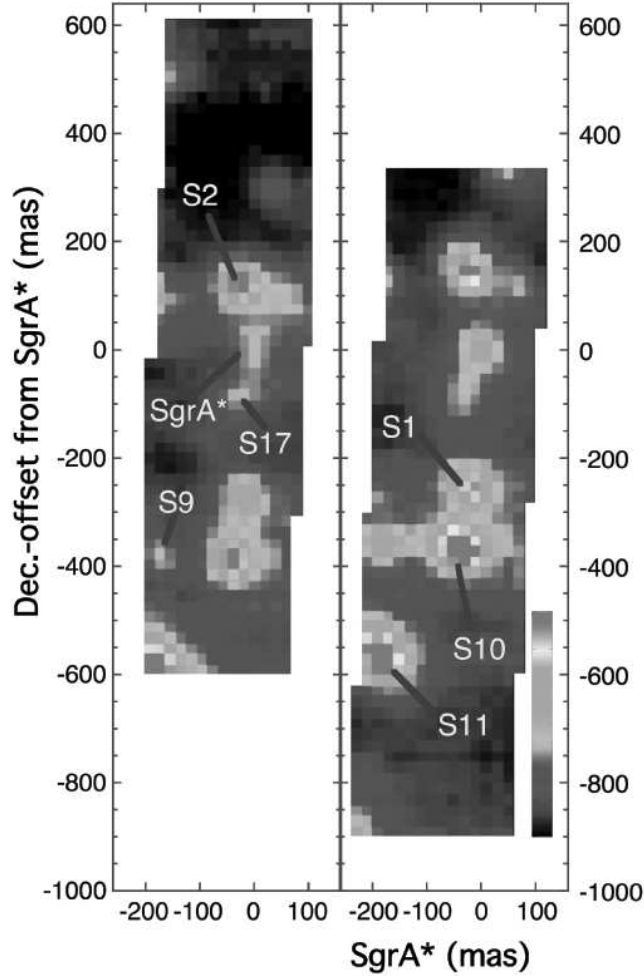}
\caption{K-band images of the immediate vicinity of SgrA*, produced by
collapsing the spectral cubes between wavelengths of 2.02
\micron\space and 2.38 \micron\space. Each frame is five minutes long
and was obtained sequentially with the Keck II LGS AO
system.  The data in the right panel are shifted 0.3\arcsec\space south
of those in the left panel. The image scale is 20 mas/pixel and the
stellar images have a FWHM of approximately 60 mas, close to the
diffraction limit of the Keck II telescope. Some of the stars have been
labeled according to \citet[][]{Eisa05}. The position of SgrA* is
indicated at (0,0) in both panels. Both frames are not
calibrated with respect to each other. However, the
flare at the location of SgrA* is clearly significantly brighter in the right
panel. \label{fig1}}
\end{figure}

\clearpage

\begin{figure}
\epsscale{0.80}
\plotone{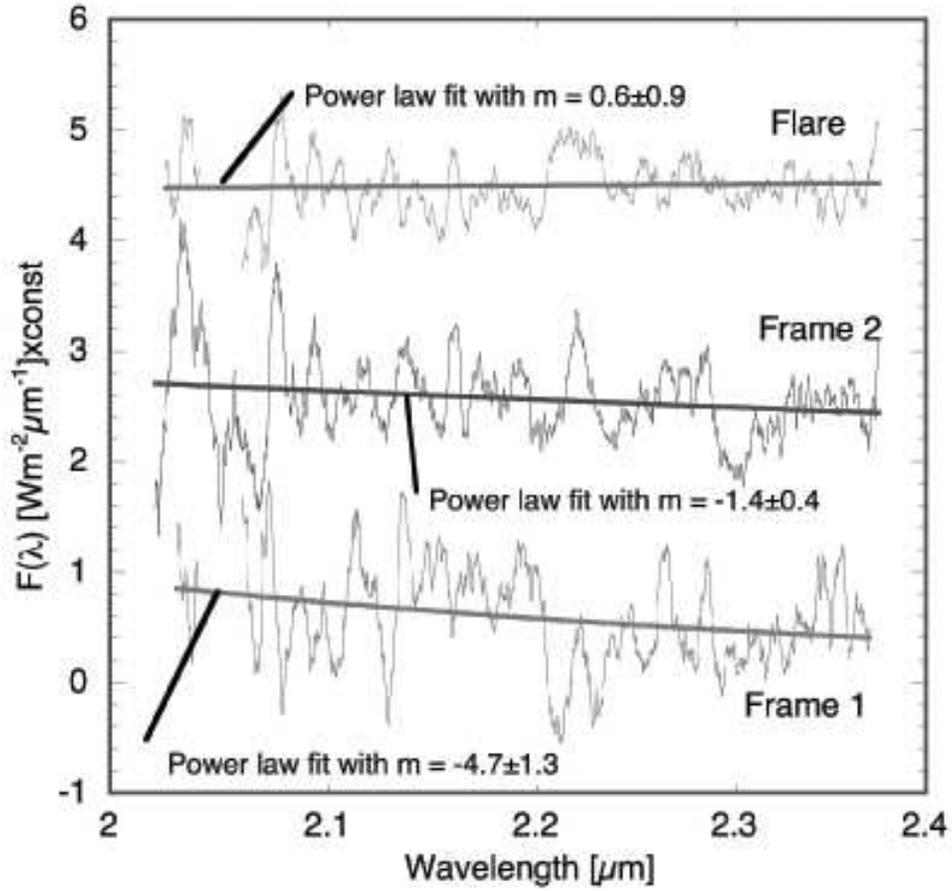}
\caption{Spectra at the location of SgrA* obtained from frame 1
(bottom) and frame 2 (middle) tied to the left axis. The middle
spectrum has been shifted up one unit for clarity. The upper spectrum
is the difference of the lower two spectra and is referred to as the
"flare" in the text. It has been shifted up by four units. The
spectra have been smoothed with a 30 pixel
wide boxcar filter for display. The lines represent power law fits to
the unsmoothed continua of the spectra. The derived values for the
slopes($F(\lambda) \propto \lambda^{m}$) are indicated.
\label{fig2}}
\end{figure}

\clearpage

\begin{figure}
\epsscale{0.80}
\plotone{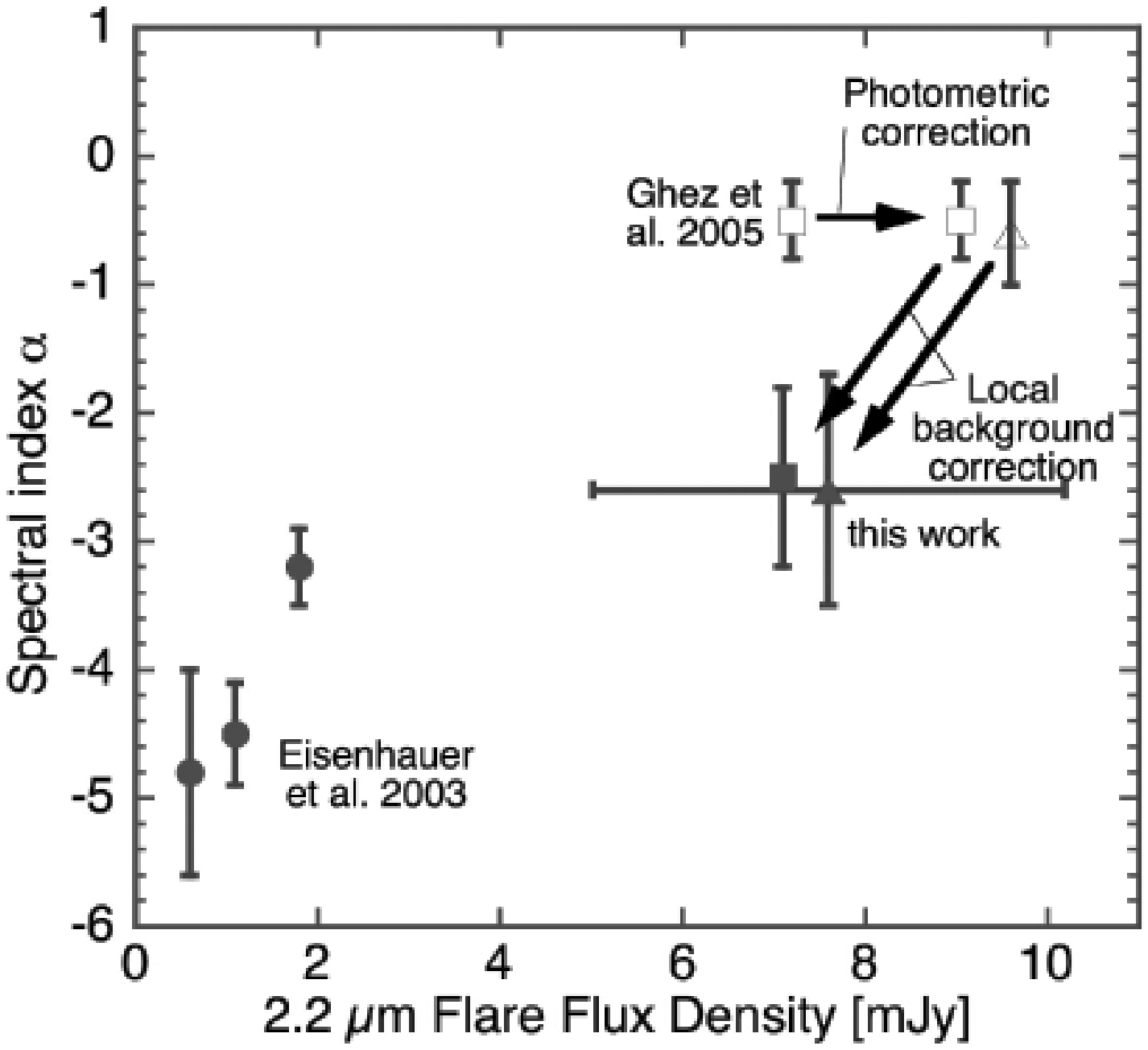}
\caption{SgrA*'s K-band spectral index ($\alpha$, where $F(\nu) \sim
\nu^{\alpha}$) as a function of its 2 \micron\space flux density.  The
results of \citet{Eisa05} and \citet{Ghea05} are indicated.  The
result of this work are denoted by the triangles.  The open triangle
marks the spectral index and flux density derived from the spectrum of
frame 2 (Fig.  2 middle), the filled triangle represents the flare
spectrum (Fig.  2 top) after subtracting the local background
(precursor) flux. The top arrow denotes the horizontal shift of the
\citet{Ghea05} data if we use their spectral index measurement to
adjust between their observed K$^{\prime}$ band to our K band (see
text) flux. Since the \citet{Ghea05} data point does not subtract a
possible precursor component, and because we do not know the
L$^{\prime}$ band brightness of the precursor, we instead use our
measured spectral shape in frame 1 in order to make a plausible
estimate of their spectral index without precursor.  The shift is
indicated by the parallel arrows. \label{fig3}}
\end{figure}

\clearpage

\end{document}